\documentclass{emulateapj}
\input{epsf.sty}

\newcommand{\dif}{\mbox{d}}

\begin{document}
\title{THE PHYSICS OF PROTOPLANETESIMAL DUST AGGLOMERATES. VI. EROSION OF LARGE AGGREGATES AS A SOURCE OF MICROMETER-SIZED PARTICLES}

\shorttitle{PROTOPLANETESIMAL DUST AGGLOMERATES VI}
\author{\scshape Rainer Schr\"apler and J\"urgen Blum}
\affil{Institut f\"ur Geophysik und extraterrestrische Physik, University of Braunschweig}
\affil{Mendelssohnstr. 3, D-38106 Braunschweig, Germany}
\email{r.schraepler@tu-bs.de}
\shortauthors{Schr\"apler \& Blum}

\slugcomment{Accepted by Astrophysical Journal}

\begin{abstract}
Observed protoplanetary disks consist of a large amount of micrometer-sized particles. \citet{DullemondDominik2005} pointed out for the first time the difficulty in explaining the strong mid-IR excess of classical T-Tauri stars without any dust-retention mechanisms. Because high relative velocities in between micrometer-sized and macroscopic particles exist in protoplanetary disks,  we present experimental results on the erosion  of macroscopic agglomerates consisting of micrometer-sized spherical particles via the impact of micrometer-sized particles. We find that after an initial phase, in which an impacting particle erodes up to 10 particles of an agglomerate, the impacting particles compress the agglomerate's surface, which partly passivates the agglomerates against erosion. Due to this effect the erosion halts within our error bars for impact velocities up to $\sim$30 m s$^{-1}$. For larger velocities, the erosion is reduced by an order of magnitude. This outcome is explained and confirmed by a numerical model. In a next step we build an analytical disk model and implement the experimentally found erosive effect. The model shows that erosion is a strong source of micrometer-sized particles in a protoplanetary disk. Finally we use the stationary solution of this model to explain the amount of micrometer-sized particles in observational infrared data of \citet{Furlan2006}.
\end{abstract}


\section{INTRODUCTION} \label{kap:INTRO}
Early theoretical investigations concerning planet formation followed two paradigms. The first paradigm considers a laminar nebula in which dust particles coagulate as a result of Brownian motion and differential drift velocities. Most particles
arriving at the midplane at 1 AU reach sizes of 1-10 cm within 1500 - 3000 orbits \citep{Weidi1980, Nakagawa1981,Nakagawa1986}. This settling leads to a dust-dominated sub disk, which keeps a finite thickness owing to shear turbulence
\citep{SchrHen04, Cuzzi1993, Weidi2006, Weidi1988, Weidi1984,Weidi1980}. The shear turbulence is generated by the difference in orbital velocity between the pressure-supported gas disk and the dust-dominated sub-disk \citep{SchrHen04, Cuzzi1993}. In the sub-disk, particles can grow to large agglomerates \citep{SchrHen04}, whereas below and above the sub-disk, a strong outward-directed wind is generated. This wind, which has typical velocities of 30 m s$^{-1}$ at 1AU \citep{Weidi2006, Cuzzi1993} carries along small particles that are perfectly
coupled to the gas.

The second paradigm considers a fully turbulent nebula. The dust within the disk coagulates due to random velocities provided by the turbulent gas motion \citep{Brauer2008,Schmitt}. The strong turbulent motion in the gas prevents the formation of a dust sub-disk. After the particles have coagulated to several decimeters in size, they begin to migrate inwards with velocities up to 50 m
s$^{-1}$, owing to friction with the sub-Keplerian gas disk \citep{WeiCu1993}.

More recently it was found that the truth may lie in a combination of both models \citep{Gammie1996, Stepinski1999,Ciesla2007}. The MHD simulations showed, that in a fully turbulent Kelvin-Helmholtz driven disk, turbulence free "dead zones" appear
at places where dust prevents the ionization of gas. In this modern paradigm of protoplanetary dust growth, fast migration of large dust aggregates as well as fast winds (above/below the sub-disk) occur at different regions in the nebula. If large dust aggregates and micrometer-sized particles coexist in the nebula, then all of the above scenarios lead to high-speed collisions between them. Thus, to understand planetesimal formation under the various scenarios described above, it is essential to understand the collision physics of the dust aggregates embedded in the protoplanetary disk. In a series of experimental papers, which are listed in  \citet{BlumWurm2008}, the growth models of coagulation by Brownian motion and relative drift motions used in \citet{WeiCu1993} and \citet{Cuzzi1993} were
basically confirmed and refined. A recent dust-growth model by \citet{Zsometal2010}, which is based upon an elaborate dust-aggregate collision model \citep{Guettleretal2010}, shows that the maximum dust aggregate size in a minimum-mass solar nebula model is around 1 cm, due to the overwhelming effect of bouncing in aggregate-aggregate collisions.

Thus,  realistic dust-growth models have two problems: (1) They fail to describe the formation of meter-sized (and larger) particles, and (2) they fail to explain the observed large amount of micrometer-sized particles in protoplanetary disks without a dust retention mechanism \citep{DullemondDominik2005}. While the latter is a real shortcoming of the models, the former might be an indication of the non-existence or short survival lifetimes of large protoplanetary dust aggregates. To overcome the depletion of small grains \citet{Okuzumi2009, Okuzumi2010a} and \citet{Okuzumi2010b} found that in case they are charged their coagulation is supressed.

In this article, we hypothesize that dust aggregates of considerable size (e.g. meter size) exist and we follow the evolution of these bodies in the protoplanetary disk. We will experimentally show that high-velocity impacts of micrometer-sized grains into large dust agglomerates are a strong source of further micrometer-sized particles, which considerably limits the size of the largest surviving bodies.

The experimental method and the laboratory setup to perform collisions between  micrometer-sized particles and dust aggregates are described in Sect. \ref{kap:EXACT}. In Sect. \ref{kap:EIMER}, the experimental results and their analysis are given, along with a numerical model that explains and confirms the experimental results. In Sect. \ref{kap:THAT}, the experimental outcomes are applied to turbulent protoplanetary $\alpha$-disks with values of the turbulence parameter $\alpha=10^{-2}$ and $\alpha=10^{-4}$, which shows the consequences of the erosional process on the population of micrometer-sized and decimeter-sized particles at different Kepler radii. In this Section, we also compare our findings to recent infrared observations of protoplanetary disks by \citet{Furlan2006}.
\section{EXPERIMENTAL APPROACH} \label{kap:EXACT}
\subsection {Mechanical Part}
To reproduce the collision scenario following the paradigms described in Sect. \ref{kap:INTRO}, which preferentially leads to collisions between the smallest and the largest constituents of the dust size distribution, we developed a laboratory experiment in which collisions between micrometer-sized dust grains (projectiles) and cm-sized porous dust aggregates (targets; simulating the surfaces of much larger bodies) can be studied. As an analog dust material for the projectiles we use spherical SiO$_2$ particles with a diameter of 1.5 $\mu$m (see \citet{BlumSchr2004}). Thus, our projectile particles have sizes typical for protoplanetary dust, consist of a silicate-like material, and have a morphology suitable for theoretical modeling. The targets were produced by random ballistic deposition (RBD) out of the same particles and have a volume filling factor of $\phi = 0.15$ \citep[see][]{BlumSchr2004,BlumSchr2006}. RBD agglomerates have the same volume filling factor (and, thus, porosity) as the ballistic particle-cluster agglomerates which should dominate the dust-aggregate structures if aggregates preferentially grow by collection of small grains.

\begin{figure}[!thb]
    \center
    \includegraphics[width=\columnwidth]{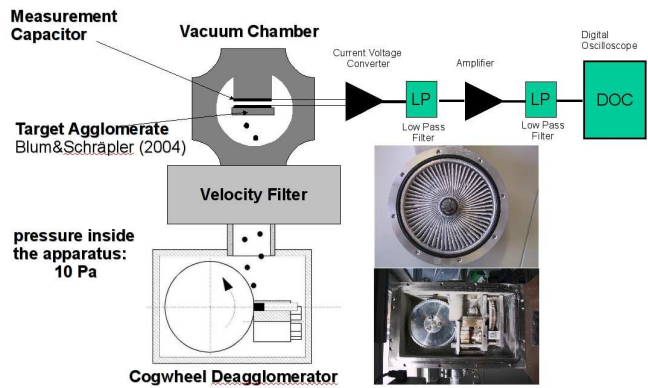}
    \caption{\label{fig:setup}Schematics of the experimental setup, consisting of a vacuum chamber, which houses the target, the
velocity filter (see upper inset photograph), the cogwheel (see lower inset
photograph), the charge-measurement capacitor, and the measurement
electronics.}
\end{figure}

To simulate the projectile-target collisions hypothesized for the protoplanetary disk, a jet of fast projectiles is directed to centimeter-sized targets in a vacuum environment. The mass loss or gain of the targets is determined by precision-weighing of the targets before and after exposure to a well-determined flux of projectiles. The experimental apparatus consists of a dust-dispersion cogwheel, a velocity filter and a vacuum chamber, which contains the target (see Fig. \ref{fig:setup}). The inside of the apparatus has a residual gas pressure of $<$4 Pa to prevent the particles from losing kinetic energy due to friction with the gas. The surfaces of the inside of the experiment are covered with vacuum grease on which the particles inevitably stick upon impact. Thus, only particles on ballistic trajectories from the cogwheel through the velocity filter can reach the target. To avoid ejected particles from falling back to the surface of the target agglomerate, the dust jet is directed in a vertically upward direction.


The particle jet is produced with a fast rotating cogwheel (see Fig. \ref{fig:setup}), which also deagglomerates the powder dust sample into single monomer grains. The cogwheel is fed with dust from 20 dust reservoirs, which are arranged on a revolver. A piston presses the dust from a chosen reservoir against the fast rotating cogwheel. The deagglomerated particles possess up to twice the circumference velocity of the cogwheel, because they nearly elastically rebound from the cogs. The small amount of particles, which are not properly deagglomerated, impact the cogwheel plastically and possess velocities up to the circumference speed of the cogwheel. Due to this difference in velocity, it is possible to separate the non-deagglomerated particles from the monomer grains by a velocity filter. The velocity filter consists of an impeller with 60 blades (see Fig. \ref{fig:setup}). The inclination of the blades is chosen such that their circumference covers 20 degrees between top and bottom of the impeller. If the impeller rotates at a constant speed, only particles within a preset velocity range can pass through the device. Particles that are either too fast or too slow hit the leading or the trailing blade of the impeller on which they stick, because the blades are covered with vacuum grease. We measured the velocity distribution of the particles leaving the filter by illuminating the particles with a pulsed laser beam and observing their trajectories by a long-distance-microscope equipped with a high-speed camera. We found that the velocity of the particles leaving the velocity filter possess a full-width-at-half-maximum (FWHM) of 17\% of the optimum velocity given by the rotational speed of the filter. The velocity filter can be operated from 0 m s$^{-1}$ to 90 m s$^{-1}$, but the cogwheel's deagglomeration efficiency drops
dramatically at rotational speeds slower than 7.5 m s$^{-1}$. To calibrate the mass flow to the target, a greased target is placed inside the apparatus instead of the agglomerate target so that impacting particles will completely stick there. Thus, the mass arriving at the target area can be measured as a function of the dust mass
provided to the cogwheel. In the second step of the measurement, the agglomerate target is weighed before and after the impact of a certain amount of particles. The strength of erosion (or growth) of the target can be determined by the ratio of the mass difference of the agglomerate before and after the exposure to the dust flux and the impacted mass. In a typical experiment, the target agglomerate is placed in the experimental chamber and the chamber is then evacuated. Upon reaching the desired vacuum, 1 to 3 dust cylinders are pressed against the rotating cogwheel. After each cylinder, the camber is carefully vented to electrically discharge the target agglomerate. Then, the agglomerate is removed from the chamber and weighed. After weighing, the target is placed back inside the experimental chamber. In typical experiments this procedure is repeated 2-10 times, depending on the experiment type.

\subsection {Electrical Part}\label{kap:elpa}
\begin{figure}[!thb]
    \center
    \includegraphics[width=\columnwidth]{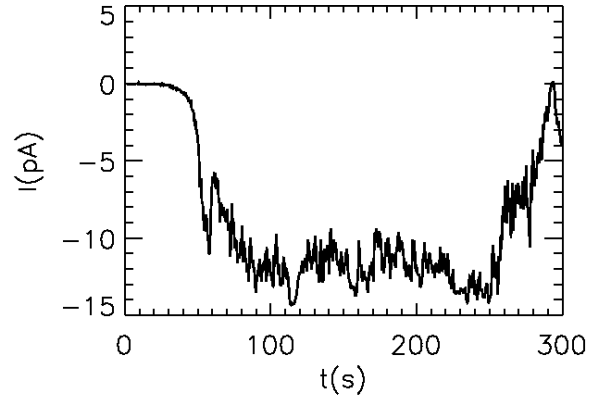}
    \caption{\label{fig:EichFett}A typical electric current profile when a greased target is exposed to a dust jet with 15 m s$^{-1}$ impact velocity. The dust flow starts at $t \approx 50$ s and ends at $t \approx 300$ s. The dust jet is generated by pressing one dust cylinder against the cogwheel.}
\end{figure}

As known from previous experiments \citep{PopSchr05, PopBH}, particles in the dust jet produced by a cogwheel are heavily charged and possess a power-law charge-velocity distribution. To assess the influence of the particle charge on our erosion measurements, we measured the charge of the impactors and the charge transfer to the target agglomerate. This was done using the effect of electrical influence. The electrical setup is also shown in Fig \ref{fig:setup}. The target agglomerate is placed on one insulated plate of a measurement capacitor. If an electric charge is deposited on the target agglomerate, an identical charge is induced on the capacitor plate, which drives an electric current to the opposing capacitor plate. This current is amplified and logged by a digital oscilloscope, which is connected to a computer. The charge on the dust agglomerate is determined by integration of the electric current over time. To measure the charge of the projectiles and compare our measurement method with those of \citet{PopSchr05}, a greased target was placed on the measurement capacitor instead of the target agglomerate. All impacting particles stick on the greased target and keep their charge. The number of elementary charges on particles,
accelerated to velocities of 15, 30, 45 and 60 m s$^{-1}$, were measured.
A typical current profile during an experiment is shown in
Fig. \ref{fig:EichFett}.


By weighing the greased target before and after the impact of the particles, the charge per particle could be determined. The average number of elementary charges on each monomer particle accelerated to velocities of 15, 30, 45 and 60 m s$^{-1}$ are 44, 198, 318, and 484 e$^-$, respectively, which is in good agreement with the measurements by \citet{PopSchr05} and \citet{PopBH}. As a result of the charged dust target, the impactors lose kinetic energy before an impact. The calculation of the energy loss due to the electric potential was approximated by assuming the target agglomerate as a charged sphere with the radius of the target diameter and the charged particles as test charges arriving from infinity. These calculations were confirmed by a numerical simulation of the incoming particles using the exact geometry of the experimental setup with the software package COMSOL. In all measured parameters, the velocity loss of the impinging projectiles was smaller than 2\%. The collision dynamics of the erosion process is not changed, because the required pull-off force of the particles is four orders of magnitude larger than the electric force close to the target surface. Charging of the particles and targets is therefore negligible for the following measurements.

\begin{figure}[!thb]
    \center
    \includegraphics[width=\columnwidth]{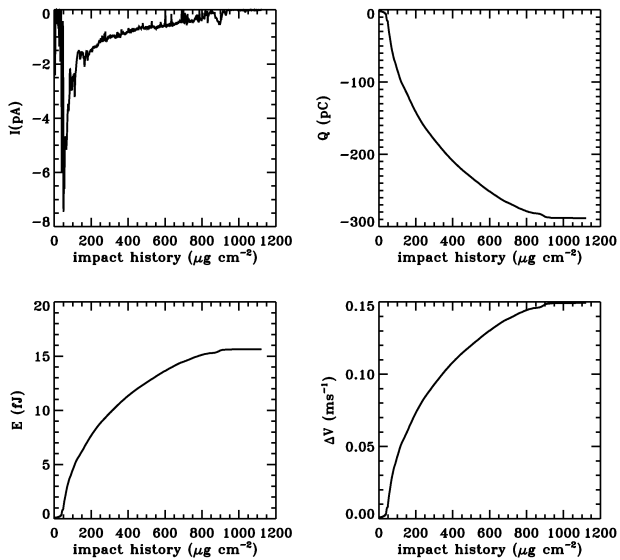}
    \caption{\label{fig:Auflad}Examples of the influence of electric charges on the collision behavior of micrometer-sized dust projectiles (with a mass of $3.5\times 10^{-15}$ kg) impacting a porous dust-aggregate target at 60 $\rm m~s^{-1}$ velocity. Top left: the profile of the electric current over the total amount of dust exposure. Top right: the total charge accumulated by the target. Lower left: the calculated energy loss of a charged projectile carrying 1000 elementary charges. Lower right: the calculated velocity loss of a charged projectile carrying 1000 elementary charges with an initial velocity of $\rm 60~m~s^{-1}$.}
\end{figure}

Fig. \ref{fig:Auflad} shows an example of the influence of the projectile and target charging on the impact energy and velocity. In the top left graph of Fig. \ref{fig:Auflad}, the electric current to the target agglomerate caused by the impinging projectiles with a mass of $3.5\times 10^{-15}$ kg is shown. The current drops to zero after a certain number of particles have impacted. This is due to the fact that particles are deposited to the surface and charges that are separated as a cause of the impact are also sitting on the surface. Each impact erodes particles from the agglomerate surface (see below) so that after a certain number of impacts, an equilibrium between deposited and separated charges (sitting on eroded particles) is reached. In the top right graph of Fig. \ref{fig:Auflad} the total charge of the target agglomerate is given as a function of the accumulated mass flux of the impinging projectiles. Note that this charge is limited by the equilibrium discussed before. In the lower left graph of Fig. \ref{fig:Auflad}, the total energy loss caused by Coulomb repulsion of an impinging particle is shown. In the lower right graph of Fig. \ref{fig:Auflad}, the velocity loss of an impinging projectile is given, which is negligible compared to the initial velocity of 60 $m~s^{-1}$.


\section{EXPERIMENTAL RESULTS} \label{kap:EIMER}
\subsection{Erosion and Passivation}
\begin{figure}[!thb]
    \center
    \includegraphics[width=\columnwidth]{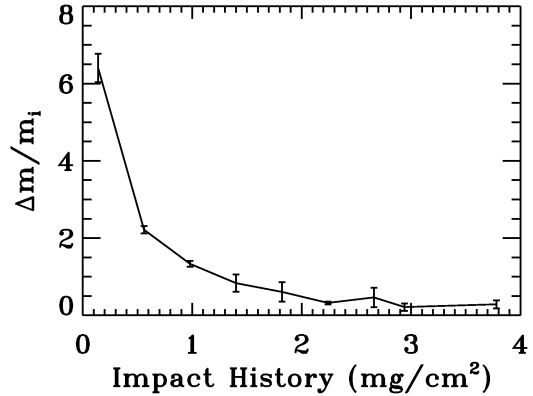}
    \caption{\label{fig:erhist}Erosion efficiency of monomer projectiles impinging a target agglomerate at 59 m s$^{-1}$. The saturation of the erosion efficiency for $\mu \stackrel{>}{\sim} 2 ~\rm mg~cm^{-2}$ is clearly visible. The error bars result from the measurement uncertainty of the weighing process.}
\end{figure}

\begin{figure}[!thb]
    \center
    \includegraphics[width=\columnwidth]{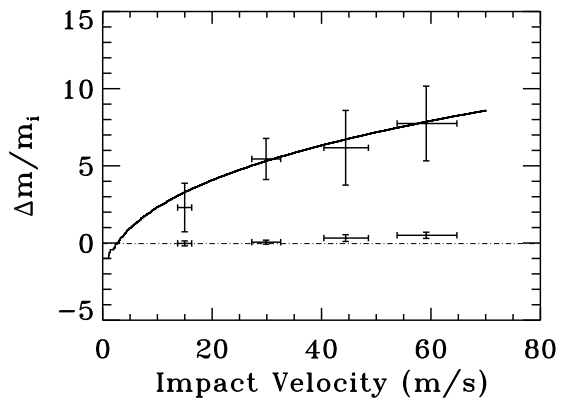}
    \caption{\label{fig:AO}Erosion efficiency of target agglomerates for impact velocities of 15, 30, 44, and 59 m~s$^{-1}$, respectively. The upper data points with error bars were recorded at the onset of the experiment, in which the target agglomerate was very porous and not yet passivated, whereas the lower data points with error bars show a reduced erosion efficiency after the target-agglomerate's surface was almost passivated against erosion. The velocity error bars denote the velocity distribution stemming from the velocity filter. The vertical error bars denote one standard deviation of several measurements with different target agglomerates. The solid curve follows Eq. \ref{eq:mv} and is the result of a numerical impact model described in Sect. \ref{sect:numunp}. }
\end{figure}

\begin{figure}[!thb]
    \center
    \includegraphics[width=\columnwidth]{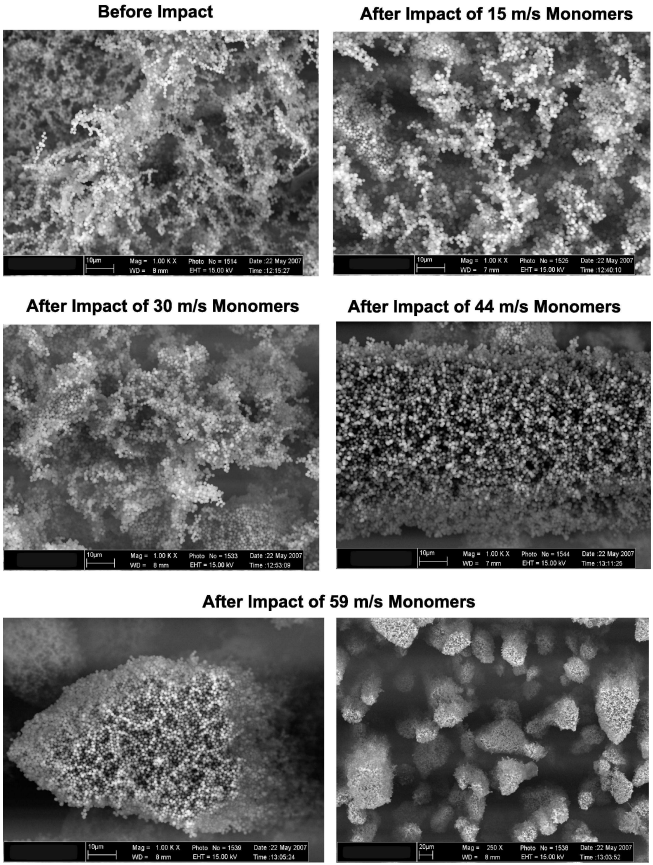}
    \caption{\label{fig:sem}SEM images of the surface of a target agglomerates before and after the passivation by dust-particle impacts at velocities of 15, 30, 44, and 59 m $s^{-1}$, respectively. The scale bars at the bottom left of each SEM image are 20 $\mu$m for the bottom-right image  -  and 10 $\mu$m  for the other images.}
\end{figure}

We performed the impact experiments described in Sect. \ref{kap:EXACT} at impact velocities of 15, 30, 44, and 59 m~s$^{-1}$, respectively. A surprising result is that the erosion efficiency (i.e. the relative mass loss of the target) depends on the impact history of the (initially highly porous) target agglomerate, as can be seen in Fig. \ref{fig:erhist} for impacts with 59 m~s$^{-1}$ projectile velocity. The data points show the mass loss of the target agglomerate, $\Delta m$, per impinging monomer mass, $m_i=3.5\times 10^{-15}$ kg, as a function of the total mass exposure per unit cross section onto the target, $\mu$. One can see from the saturation of the curve that the target agglomerate is passivated against erosion for mass exposures $\mu \stackrel{>}{\sim} 2 ~\rm mg~cm^{-2}$. This saturation mass exposure is about 20 times the number of particles on the surface layer of an unprocessed agglomerate. Experiments with impact velocities of 15, 30, 44, and 59 m s$^{-1}$ show a similar saturation behavior. The error bars in Fig. \ref{fig:erhist} denote the measurement uncertainties of the weighing process. The erosion efficiency at the initial and final stage for all measured impact velocities is shown in Fig. \ref{fig:AO}. The velocity error bars denote the velocity dispersion due to the width of the velocity filter. The error bars in the vertical direction are the standard deviation of several measurements with different target agglomerates. Fig. \ref{fig:AO} shows that the initial erosion efficiency scales with impact velocity but is reduced to a saturation level of zero for impact velocities lower than 26 m s$^{-1}$. For higher impact velocities, the erosion efficiency is reduced to about 15\% of its initial value. The reduction of the erosive effect of the particle impacts goes along with a compaction of the surface. Figure \ref{fig:sem} shows scanning electron microscopy (SEM) images of the target agglomerate's surface before and after an exposure to dust fluxes at 15, 30, 44, and 59 $\rm m~s^{-1}$, respectively. An obvious change in surface structure occurs between 30 and 44 $\rm m~s^{-1}$.


\subsection{Theoretical Considerations} \label{kap:theco}
\subsubsection{Analytical Part}
A key to understanding the erosion behavior described above is given in the work of \citet{Konsti1999}. He numerically simulated the formation of agglomerates by the RBD process and discovered that the coordination number $n$ (i.e. the number of next neighbors) in an agglomerate increases with increasing impact velocity of the depositing particles, until a maximum coordination number of $n_{\rm max}=6$ is reached. Note that the maximum coordination number found by \citet{Konsti1999} has a gaussian distribution with a FWHM of 2. \citet{Konsti1999} found that a relation fits his numerical results which reduces the collision of a single particle with a particle bound in a large agglomerate to a collision of a single particle with another single particle of an effective mass given by
\begin{eqnarray}
m_{\rm eff}=m(1+C n)\mbox{,}  \label{eq:effm}
\end{eqnarray}
where $C$ is a dimensionless deposit ``rigidity'' parameter in the range $(1,\infty)$,  $m$ is the mass of the particle bound to the agglomerate and $n$ its local coordination number. Using Eq. \ref{eq:effm} and applying momentum and energy equations we calculated the energy transferred to an effective particle to
\begin{eqnarray}
E_{\rm eff}=4 E_{imp}\frac{1+C n}{(2+C n)^2} \mbox{,}\label{eq:Ee}
\end{eqnarray}
with $E_{imp}$ denoting the kinetic energy of the impactor. The energy to release a particle from an agglomerate is given by
\begin{eqnarray}
E_{r}=E_{c}~n\mbox{,} \label{eq:Er}
\end{eqnarray}
where $E_c$ is the break-up energy of a single particle-particle contact. An empirical relation for the capture threshold velocity $v_{st}$ can be deduced from the experiments by \citet{PoppeBlumHenning2000}, who showed that
\begin{eqnarray}
v_{st}=0.92\left(\frac{a_0}{\mbox{$\rm \mu m$}}\right)^{-0.52}
\mbox{m s}^{-1}\mbox{,}
\end{eqnarray}
with $a_0$ being the monomer radius. Based on this relation, we can derive the break-up energy of a particle contact to be
\begin{eqnarray}
E_{c}=\left(\frac12\right)^{\frac73} m_0 v_{st}^2  \mbox{,} \label{eq:Ec}
\end{eqnarray}
\citep{BlumWurm2000} with $m_0$ being the mass of the impacting monomer. It is obvious from Fig. \ref{fig:sem} that the target agglomerate gets compacted by the impinging dust projectiles. An increased compression, i.e. an increased volume filling factor $\phi$ results in an increased coordination number \citep{vandelagemaatetal2001}. The increased coordination number increases
the release energy of a particle and is a mechanism that passivates the agglomerate.

\subsubsection{\label{sect:numunp}Numerical Description of an Unpassivated Target Agglomerate}
\begin{figure}[!thb]
    \center
    \includegraphics[width=\columnwidth]{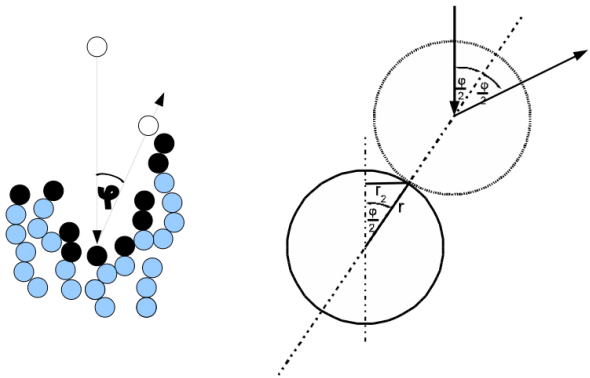}
    \caption{\label{fig:phi}Left picture: An impacting particle (white) is shown that is reflected off an agglomerates surface at an angle $\varphi$. The black particles are surface particles that can be reached by the impinging particle. The grey particles are bulk particles of the agglomerate that cannot be reached. The angle $\varphi$ corresponds to the probability that the impactor can impact a second time. Right picture: Relation between the maximum escape angle and the probability of a second impact of the impinging particle $r_2 r^{-1}$.}
\end{figure}

\cite{DominikTielens1997} and \cite {WadaII} predict an erosion effect proportional to the impact energy, thus proportional to the squared impact velocity, which is in contradiction to our experimental results (see Fig. \ref{fig:AO}).
To validate our measurements by theory, we developed a numerical model that applies the collision model of \citet{Chokshi} with the adaptation recommended by \citet{BlumWurm2000}. In our model, the energy loss of a repelling particle equals its binding energy and the effective mass term (Eq.~\ref{eq:effm}) of \citet{Konsti1999} is used. The effective mass term requires a rigidity parameter $C$. For fluffy agglomerates \citet{Konsti1999} chose it close to 1 to get realistic results. Therefore we also chose a value 1 for the rigidity parameter. In our model, all particles of an agglomerate have an effective mass corresponding to their coordination number and thus only part of the impactor's kinetic energy is transferred to the respective agglomerate monomer. The remaining kinetic energy of the impactor after its impact is either used in a further impact or is lost if the particle escapes from the agglomerate. The escape probability ${\Xi}$ of the impactor and the ejected particles is calculated from the surface texture of a numerically grown RBD agglomerate \citep[see][]{BlumSchr2006}. This calculation is done in three steps. In the first step, all particles that can be reached by impinging particles from a direction perpendicular to the (mean) target surface are identified and defined as surface particles (see Fig. \ref{fig:phi}). Depending on the fraction of the particles on the aggregate's surface that can be reached by impinging particles of the same size, a corresponding accessability is calculated.  In a second step, the angles of the surface particles are calculated under which they can escape without impacting again with the target agglomerate after they are released by the original projectile. The same is done for the reflected impactor.  For simplicity this procedure is not applied to all particles and neighbor particles on the agglomerate surface. The problem was reduced to a 2D-problem.

Only particles on a given line on the surface are taken into account, and to calculate the required angles, only particles are used that are found in one direction of that line. It is therefore assumed that the target surface is isotropic. The smallest of  the angles from a given particle on the line to its neighbors in one direction is the maximum escape angle (see Fig. \ref{fig:phi}). By repeating this procedure for a large number of particles on the defined line on the agglomerate's surface and by averaging over the occurring angles, the mean escape angle is determined.


 The probability to impact a further particle is calculated by assuming that this is the case if the reflection angle is larger than the mean escape angle. The probability of a particle to fulfill this condtion  is found by taking into account that all particles are spherical. The reflection angle of a particle impinging perpendicular to the agglomerates surface is connected with  the impact distance from the target particle's axis $r_2$ in the agglomerate (see Fig. \ref{fig:phi}).

The correspondence between the reflection angle and the probability of an reflection angle larger than $\phi$ is given by
\begin{eqnarray}
\Xi=\frac{r-r_2}{r}
\end{eqnarray}
which corresponds to
\begin{eqnarray}
\Xi= 1-\sin\frac{\phi}{2}
\end{eqnarray}
(see Fig. \ref{fig:phi}). These probabilities are multiplied with the corresponding accessibility of that given particle and averaged over all considered particles. We found that for RBD agglomerates ${\Xi}=0.39$.

If the energy transferred to the agglomerate target exceeds the binding energy of the monomer grain, it will be ejected and will impact another agglomerate monomer with a probability of $1-{\Xi}$. This corresponds to the fact that the mean coordination number of the (initial) target agglomerate is 2, which means that it is an agglomerate of particle chains. It is very unlikely for an impactor to hit the end of a chain in a direction parallel to the chain. In all the other impacts the hit particle can be directly ejected. This is not the case for a compacted agglomerate with a larger mean coordination number.

The contact energy $E_c$ of a monomer particles required for the modified \citet{Chokshi} model (see also \citealp{DominikTielens1997}) is taken from \citet{PoppeBlumHenning2000}. The coordination number distribution of the particles at the agglomerate's surface and the surface texture of the agglomerate are also calculated from the numerically grown RBD agglomerate \citep{BlumSchr2006}. Our numerical erosion model is realized as a statistical code with recursive elements. This means that all possible impacts (concerning the effective mass of the target particles) of an impinging particle are calculated, multiplied with the probability to release a particle and summed up. All possible impacts of the reflected impactor and the ejected particles are multiplied with the probability of a further impact and the probability of the occurrence of the effective masses of the targets and are also summed if their impact energy is large enough to release a particle. This is done in a recursive way until the particle energy of a given branch of the simulation is lower than $E_c$ which then leads to sticking. In other words, to get the total number of particles that are ejected by the impact of a given particle, the probabilities of all possibilities to eject a particle are summed up, where the possibility to eject a certain particle is the product of all possibilities of the preceding path of the impactor. The outcome of the impact simulation on the unprocessed agglomerate is shown as a solid curve in Fig. \ref{fig:AO}. It can be fitted for impact velocities larger than 15 m~s$^{-1}$ by a squareroot function
\begin{eqnarray}
\frac{\Delta m}{m_i} = \sqrt{\frac{1.25~v_{imp}}{1~\mbox{m s}^{-1}} - 7.5} \label{eq:mv}.
\end{eqnarray}
In the region lower than 15 m~s$^{-1}$ sticking can occur, which considerably reduces the erosion efficiency. For velocities below 2.7 m~s$^{-1}$, sticking dominates the impact process, which leads to a net growth of the agglomerate i.e. $\Delta m /{m_i} < 0$. At velocities lower than 1 m~s$^{-1}$ all impactors stick i.e.  $\Delta m /{m_i} = -1$.

\subsubsection{Numerical Description of a Passivated Target Agglomerate}\label{sect:NDPA}

In a passivated agglomerate, the projectiles impact perpendicular to the surface of interconnected particles. The ejection of such a particle requires more energy than for the loosely-bound monomers in the RBD targets. As no numerical agglomerate for passivated targets is available, we have to rely on the structural information provided by the SEM pictures, which are two-dimensional. Therefore it is not possible to exactly measure the coordination number of a passivated agglomerate. We  also choose C=1 for this simulation. The coordination number of 6 found by \citet{Konsti1999} leads to a too perfect passivation at nearly all velocities. Koji Wada (personal communication 2009) found in molecular-dynamics calculations that agglomerates that are processed by collisions tend toward a mean coordination number of 4. Therefore we chose a mean coordination number distribution on the agglomerate surface of 3 with a gaussian distribution of a FWHM of 2. The passivated agglomerates show a steep valley-hill-structure (see Fig. \ref{fig:sem} bottom right).

\begin{figure}[!thb]
    \center
    \includegraphics[width=\columnwidth]{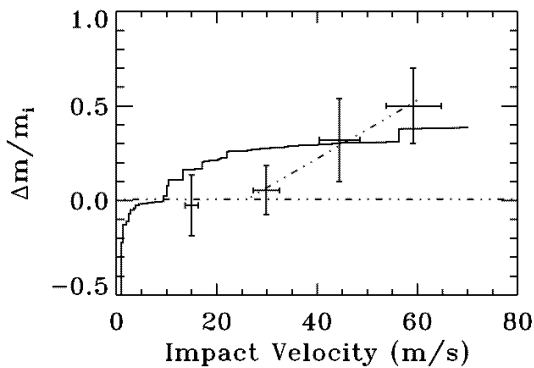}
    \caption{\label{fig:pas} Erosion efficiency as a function of impact velocity for a passivated agglomerate target. The crosses denote the experimental values with error bars, the solid line shows the result of our numerical model. The dot dot dashed lines mark the zero erosion line and the fit for the velocity dependent erosion given in Eq.~(\ref{eq:fit}) and used in Sect. \ref{kap:THAT}.}
\end{figure}

SEM images of the type shown in Fig. \ref{fig:sem} were used to estimate the probability $\Xi_p$ that an ejected or reflected particle will again impact at the agglomerate. In a passivated agglomerate, the monomers bound in the agglomerate cannot be directly ejected as it was the case with the unpassivated agglomerates, which mainly consists of particle chains. After an impactor hits a monomer bound in a passivated agglomerate, the bound monomer itself impacts its up to 5 neighbors. If its kinetic energy after this multiple impact is higher than the binding energy to its neighbors, the monomer is ejected. In our numerical code, all possible impacts with respect to the coordination number distribution were calculated and multiplied with the estimated occurrence of each coordination number. Any target monomer that is hit by any other monomer particle (projectile or ejected target monomer) can again impact with its neighbor particles in a recursive way. All ejected particles can again impact on the agglomerate with the probability of $\Xi_p$. Please note that because of the lacking information on the internal structure of the passivated agglomerate, this numerical method can only be a rough estimate. As a result of our crude model for passivated targets (see Fig. \ref {fig:pas}), it was found that the order of magnitude of the agglomerate passivation against erosion can be reproduced by the numerical model. However, due to the crude assumptions of our model, it is not able to reproduce the detailed characteristics of the erosion efficiency as a function of impact velocity. Thus, for the following application (Sect. \ref{kap:THAT}), the measured values are used via a linear interpolation function of the measured data
\begin{equation}
\frac{\Delta m}{m_i}=a v_{imp} + b \label{eq:fit}
\end{equation}
with $a = 0.0155~ \rm s~m^{-1}$ and $b = -0.4$ in the velocity range $26 {\rm m~s^{-1}} \le v \le 60 {\rm m~s^{-1}}$.


\section{APPLICATION TO A PROTOPLANETARY DISK} \label{kap:THAT}
\subsection{Grain Population in a Solar Nebula}
To estimate the influence of the erosion process on the dust-aggregate distribution in a protoplanetary nebula, a simple analytical model based on the timescales of the erosion and coagulation processes will be developed and compared to observational findings in the following section. For this we use a standard minimum mass protoplanetary nebula with a disk mass of 0.0425 M$_\odot$. The analytical disk model is taken from \citet{SchrHen04} (their Appendix A1). This leads to a surface mass density of the  at 1 AU of $1.7\times 10^4$ kg~m$^{-2}$ and a gas density in the midplane at 1AU of $1.4 \times 10^{-6}$ kg m$^{-3}$ and 280 K at 1 AU. The dust mass is 1\% of the gas mass. The relative velocities of the particles are taken from \citet{Weidi77} (his Figures 3 and 5) and approximated and extrapolated in Eqs. \ref{eq:vil}, \ref{eq:radv} and \ref{eq:AUv}). The number of large particles $n_m$ is calculated for a given Kepler radius by assuming that 70\% of the total mass is in the large dust agglomerates with a size given in Figs.~\ref{fig:ET-2} and \ref{fig:ET-4} (bottom graph, solid line). The total mass ratio of 70\% for the large dust aggregates is taken from \citet{Nomura2006} (their Fig. 5) after one coagulation timescale, the radii of the large dust aggregates are taken from \citet{Nomura2006} (their Fig. 6, dash-dotted line). A increase of the total mass in the large agglomerates to 90\% of the total mass, which would be the case after 2 coagulation timescales in \citet{Nomura2006} (their Fig. 5) leads to minor changes of our results that wuold be  hardly visible in  Figs.~ \ref{fig:ET-2} and \ref{fig:ET-4}. We ignored icy grains at outer disk radii. The threshold impact velocity for erosion should be considerabely higer for icy particles. However the predominent part of the erosion takes place at Kepler radii below 2AU.

\subsubsection{The Erosive Formation of Monomer Dust Grains and Their Subsequent Coagulation}\label{kap:TEF}
\begin{figure}[!thb]
    \center
    \includegraphics[width=\columnwidth]{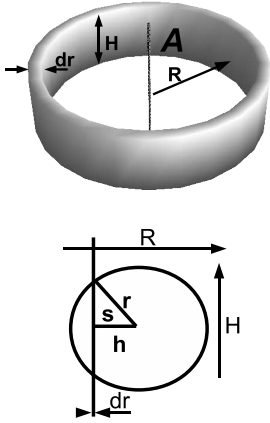}
    \caption{\label{fig:da}Upper picture: The differential volume element  $\dif V$, which is a torus in the protoplanetary nebula at the orbital distance $R$ with a height $H$ and a thickness of $\dif r$. Lower picture: A large particle with a radius $r$ at a radial distance $s$ from $\dif V$ in the midplane of the disk. As $s \le r$ an area of intersection forms as a ring with a radius $s$ and a thickness $\dif r$.}
\end{figure}

Most of the coagulation models \citep[e.g.][]{Nomura2006} show that micrometer-sized particles vanish from the nebula on a very short timescale, due to their effective coagulation. Therefore, it is important to calculate whether the source of small dust grains released by the erosion of large dust aggregates, as described in the previous sections, is strong enough to compensate the loss of the dust particles by coagulation. As we have shown above, a small initial number of micrometer-sized dust grains is sufficient to start the erosion process of aggregates with sizes of decimeter and above. The particle production rate in a differential volume element $\dif V=2 \pi R H \dif r$ (see Fig. \ref{fig:da}) is given by
\begin{eqnarray}
\frac{\dif(\dif N)}{dt}=\rho_{0} \  n \  v_{imp} \  f \  { \dif A_m} \mbox{,}\label{eq:er}
\end{eqnarray}
where the relative velocity between the micrometer-sized particles and the large agglomerates is taken from  \citet{Weidi77} and approximated for Kepler radii larger than 0.2AU  to
\begin{eqnarray}
v_{imp}= 50 \rm{\frac ms} R[AU]^{-\frac13} \mbox{.} \label{eq:vil}
\end{eqnarray}
Here, $f(v_{imp})$ is the number of monomer grains ejected by an impacting particle, $\rho_0$ is the number density of micrometer-sized particles in the midplane of a protoplanetary disk, and $n$ is the number of large dust aggregates at a disk area $A_D=2 \pi R r$, where it is implicitly assumed that most of these large aggregates are located close to the midplane. This area element is chosen because all particles located in there cut the differential volume element $\dif V$ with their front side in the velocity direction. Therefore the number $n$ can be expressed by the surface density $\delta$ of large agglomerates
\begin{eqnarray}
n=\delta 2 \pi R r \mbox{.} \label{eq:dd}
\end{eqnarray}

The mean differential area with which the particles inside $A_D$ cut $\dif V$, is given by
\begin{eqnarray}
{\dif A_m}&=&2\pi \frac{\dif r}{r} \int^0_r { s \ ds}\mbox{.}
\end{eqnarray}
With
\begin{eqnarray}
 s&=&\sqrt{r^2-h^2} \mbox{  and  } \dif s=-\frac{h \dif h}{\sqrt{r^2-h^2}}
\end{eqnarray}
we get
\begin{eqnarray}
{\dif A_m}&=&\pi r \dif r \mbox{,}\label{eq:dam}
\end{eqnarray}
assuming that all particles are homogeneously distributed in $A_D$. Here, $h$ is the distance of the center of a given particle to $\dif V$ (see Fig. \ref{fig:da}).

The number of micrometer-sized particles produced within a given radius element $2\pi R \dif r$ is $\dif N$, which will in the following be expressed using the corresponding particle number density in the midplane $\rho_{0}$. The differential surface density of the particle number $\dif N$ at a given radius element $2\pi R \dif r$ is
\begin{eqnarray}
\Sigma=\frac{\dif N}{2\pi R \dif r} \mbox{.}\label{eq:Sig}
\end{eqnarray}
The integration
\begin{eqnarray}
\Sigma=\int_{-\infty}^\infty {\rho(z) \dif z}
\end{eqnarray}
of the vertical number density structure
\begin{eqnarray}
\rho(z)=\rho_0 \exp\left(-\frac{z^2}{2H^2}\right)\label{eq:dhf}
\end{eqnarray}
in the protoplanetary nebula results in
\begin{eqnarray}
\rho_{0}= \frac{\Sigma}{H \sqrt{2\pi}} \mbox{.}\label{eq:rosi}
\end{eqnarray}
Inserting Eq. \ref{eq:Sig} yields
\begin{eqnarray}
\dif N=\rho_0  (2\pi)^{\frac32} R H \dif r \label{eq:NR1}
\end{eqnarray}
and
\begin{eqnarray}
\frac{\dif (\dif N)}{\dif t}=\frac{\dif \rho_0}{\dif t}  (2\pi)^{\frac32} R H \dif r \mbox{.}\label{eq:NR}
\end{eqnarray}

Inserting Eqs. \ref{eq:NR}, \ref{eq:dd}, and \ref{eq:dam} into Eq. \ref{eq:er} results in
\begin{eqnarray}
\frac{\dif \rho_0}{\dif t}=\rho_0 \frac{r^2 \delta v_{imp} f \sqrt{\pi}}{\sqrt{2}H} \mbox{,} \label{eq:der}
\end{eqnarray}
which is the differential equation for the dust number density increase of micrometer-sized particles without coagulation.

In the following step we will introduce the coagulation of the $\rm \mu m$-sized particles. For simplicity and as a lower limit to the erosion efficiency we assume that all micrometer-sized particles that coagulate to dimers or larger aggregates are lost to the erosion process. This assumption leads to the equation
\begin{eqnarray}
\frac{\dif \rho}{\dif t}=-\sigma v_B \rho^2 \label{eq:drho}
\end{eqnarray}
\citep{Blum}, where we assumed that the main source for the inter-particle collisions is the Brownian motion with a mean collision velocity $v_B$ and $\sigma$ is the cross section in the collision of two micrometer-sized particles and $\rho$ is given by Eq. \ref{eq:dhf}. Because $\rm \mu$m-sized particles are perfectly coupled to the disk gas, turbulence effects are negligible. To calculate the coagulation particle losses we have to take into account that the coagulation efficiency depends on $\rho$ and therefore on the disk scale height $H$. Because dust particles of $\mu$m size are mixed over $H$ on a timescale of an orbital period it is possible to integrate both sides of Eq. \ref{eq:drho}  over the disk height following \citet{Brauer2008} (in their Appendix B) which results in
\begin{eqnarray}
\frac{\dif\Sigma}{\dif t}=-\sqrt{\pi} H \sigma v_B \rho_0^2 .
\end{eqnarray}
With Eq.~\ref{eq:rosi} we obtain an equation for the midplane number density of monomers
\begin{eqnarray}
\frac{\dif \rho_0}{\dif t}=-\frac{\sigma v_B \rho_0^2}{\sqrt{2}}\label{eq:siro}
\end{eqnarray}
Performing the coagulation feedback by adding Eq.~\ref{eq:siro} to the righthand side of Eq. \ref{eq:der} results in
\begin{eqnarray}
\frac{\dif \rho_0}{\dif t}&=&\Theta \rho_0 -\Phi \rho_0^2 \label{eq:dglc}
\end{eqnarray}
 with
\begin{eqnarray}
\Theta&= &  \frac{r^2 \delta v_{imp} f \sqrt{\pi}}{\sqrt{2}H}  \\
\Phi&=&\frac{\sigma v_B}{\sqrt2}
\end{eqnarray}
whose solution is
\begin{eqnarray}
\rho_0(t)&=&\rho_{00}\frac{\Theta \exp(\Theta t)}{\Theta+\rho_{00}\Phi(\exp\Theta t -1)} \mbox{.} \label{eq:rho0e}
\end{eqnarray}
Here, $\rho_{00}$ is the initial number density of micrometer-sized dust. The half-value time of the above equation,
\begin{eqnarray}
t_{2}=\frac1{\Theta} \ln\left(\frac{\rho_{00}\Phi-\Theta}{\rho_{00}\Phi-0.5\Theta}\right),
\end{eqnarray}
seems to be dependent on $\rho_{00}$, but for a typical protoplanetary nebula $\Theta \gg \rho_{00} \Phi$, which is valid for any reasonable $\rho_{00}$. Therefore, the half-value time is
\begin{eqnarray}
t_{2}\approx \frac1{\Theta} \ln 2\mbox{.}
\end{eqnarray}
For a minimum mass solar nebula at 1 AU $t_{2}$ is about 2 years. Because $H$ is roughly proportional to $R$, $t_{2}$ approximately scales with the Kepler radius squared. As we will see later, it is of the same order as the timescale for erosional radius loss of large particles. Therefore, the stationary solution of Eq. \ref{eq:dglc} is used in the following as a zero-order approximation, i.e.
\begin{eqnarray}
\rho_{0s}=\frac{r^2 \delta v_{imp} f \sqrt{\pi}}{ H \sigma v_B } \mbox{.} \label{eq:rhor}
\end{eqnarray}

\subsubsection{Timescale of the Erosive Radius Loss of Large Dust Aggregates}
The number of the monomers eroded from a large dust aggregate per unit time is given by
\begin{eqnarray}
\frac{\dif e}{\dif t}&=&-r^2\pi v_{imp} f \rho_{0s}\mbox{.} \label{eq:ero}
\end{eqnarray}
The relation between the radius of a large particle $r$ and the number of monomer grains $e$ inside the aggregate is
\begin{eqnarray}
e=\frac{\frac43 r^3 \pi l }{V_{\mu}}~.
\end{eqnarray}
Here, $l$ is the filling factor of the large dust aggregate and $V_{\mu}$ denotes the volume of a micrometer-sized particle. Differentiation by $t$ gives
\begin{eqnarray}
\frac{\dif e}{\dif t}&=&\frac{4 r^2 \pi l}{V_{\mu}}\frac{\dif r}{\dif t}\mbox{.}\label{eq:rade}
\end{eqnarray}
Combining Eqs. \ref{eq:ero} and \ref{eq:rade} results in
\begin{eqnarray}
\frac{\dif r}{\dif t}=-\frac{1}{4 l} v_{imp} f \rho_{0s} V_{\mu}\label{eq:rl}\mbox{.}
\end{eqnarray}

\begin{figure}[!thb]
    \center
    \includegraphics[width=0.4\textwidth]{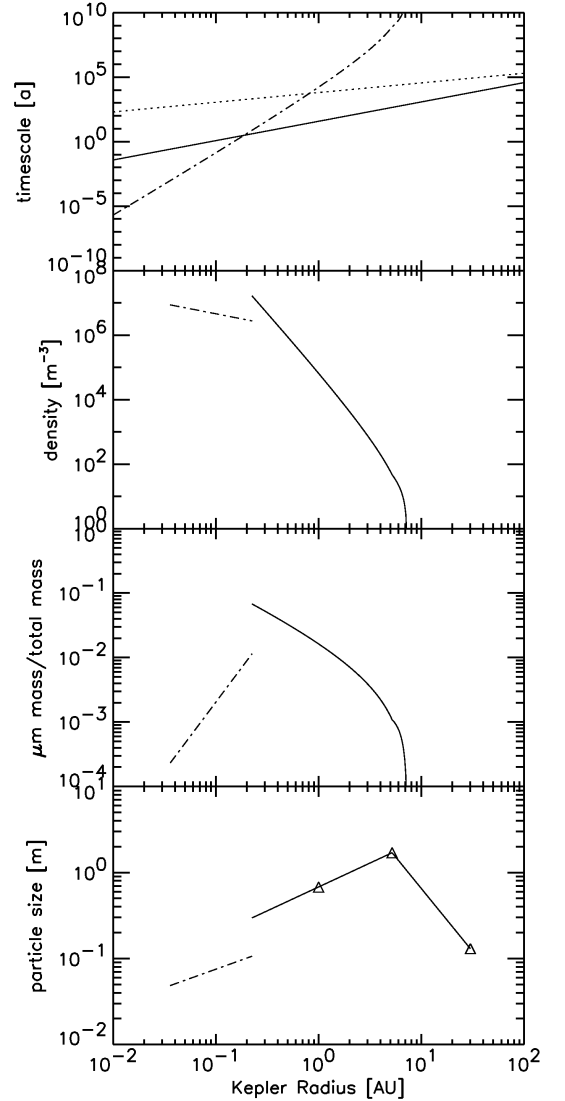}
    \caption{\label{fig:ET-2}Protoplanetary disk with $\alpha=10^{-2}$. Top graph: coagulation timescale (solid line), erosional timescale (dashed-dotted  line), radial turbulent transport (dotted line). Second  graph: number density of micrometer-sized particles at erosion timescale larger than the coagulation timescale (solid line), number density of micrometer-sized particles at erosion timescale smaller than the coagulation timescale (dashed-dotted line). Third graph: ratio of the total mass of micrometer-sized particles to the total dust mass at a given Kepler radius. Bottom graph: Particle size in which 70\% of the total solid disk mass is contained (also maximum particle size) for erosion timescales larger than the coagulation timescale taken from \citet{Nomura2006} (triangles) and their power-law interpolation (solid line). Maximum particle size for erosion timescales smaller than the coagulation timescale (dashed-dotted line).}
\end{figure}

\begin{figure}[!thb]
    \center
    \includegraphics[width=0.4\textwidth]{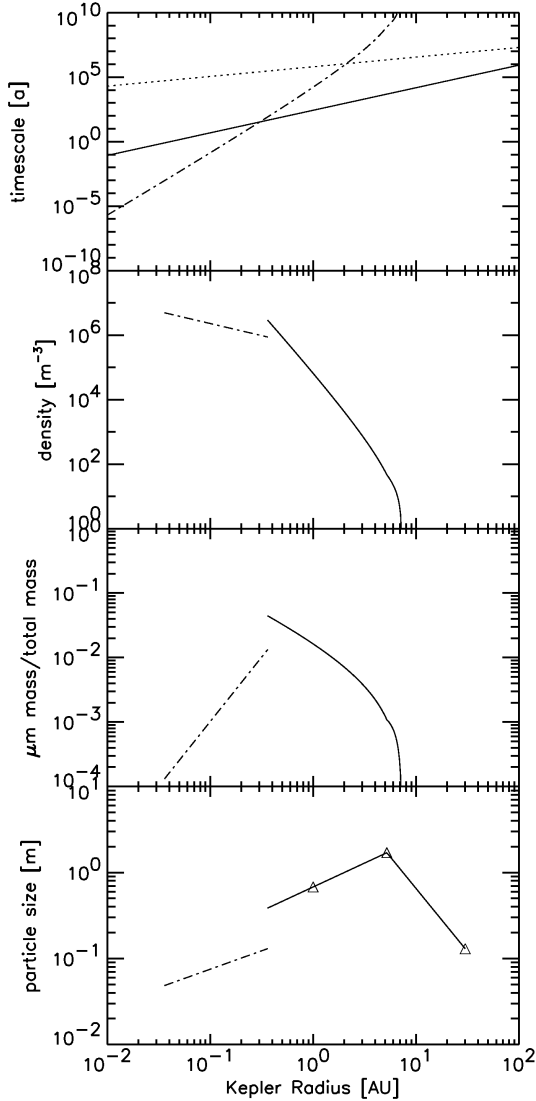}
        \caption{\label{fig:ET-4}Protoplanetary disk with $\alpha=10^{-4}$. Top graph: coagulation timescale (solid line), erosional timescale (dashed-dotted  line), radial turbulent transport (dotted line). Second  graph: number density of micrometer-sized particles at erosion timescale larger than the coagulation timescale (solid line), number density of micrometer-sized particles at erosion timescale smaller than the coagulation timescale (dashed-dotted line). Third graph: ratio of the total mass of micrometer-sized particles to the total dust mass at a given Kepler radius. Bottom graph: Particle size in which 70\% of the total solid disk mass is contained (also maximum particle size) for erosion timescales larger than the coagulation timescale taken from \citet{Nomura2006} (triangles) and their power-law interpolation (solid line). Maximum particle size for erosion timescales smaller than the coagulation timescale (dashed-dotted line).}
\end{figure}

The value $v_{imp}$ does not depend on $r$ for particle sizes larger than 0.3 m  in \citet{Weidi77} (his Fig. 2). Due to an expected filling factor of 0.4 for macroscopic protoplanetary dust aggregates and a bulk density of 2000~kg m$^{-3}$ (see \citep{Zsometal2010}), which leads to an aggregate density of 800 kg m$^{-3}$, we re-calculated the relative velocities of \citet{Weidi77} for aggregates of 40\% filling factor. For these dust aggregates, the velocity does not depend on the
aggregate size for radii larger than 0.6 m at 1AU, which is in zeroth order approximation also the case for larger Kepler radii.
with $r_0$=1m. Inserting Eq.~\ref{eq:rhor} into Eq.~\ref{eq:rl} gives
\begin{eqnarray}
\frac{\dif r}{\dif t}=-\frac{v_{imp}^2 f^2 V_{\mu}\delta r^2\sqrt{\pi}}{4 l H \sigma v_B}
\end{eqnarray}
whose solution is
\begin{eqnarray}
r=\left(\frac{1}{r_0}+\frac{v_{imp}^2 f^2 V_{\mu}\delta \sqrt{\pi}}{4 l H \sigma v_B} t \right)^{-1}
\end{eqnarray}
The  corresponding half value time  of $r$ is
\begin{eqnarray}
t_{\frac12}= \frac{4 l H \sigma v_B}{v_{imp}^2 f^2 V_ {\mu}\delta r_0\sqrt{\pi}}
\end{eqnarray}
The erosion timescale as a function of the Kepler radius is given in Figs. \ref{fig:ET-2} and \ref{fig:ET-4} (top graphs, dashed-dotted lines) for turbulence parameters $\alpha=10^{-2}$ and $\alpha=10^{-4}$, respectively. The values for $f$ for different velocities were taken from the data points of Fig. \ref{fig:pas}, where it was assumed that the curves in between the data points are straight lines.


\subsubsection{Passivation Timescale}
To decide whether the erosion efficiencies of the unpassivated or the passivated agglomerates should be used to describe the erosion processes in the solar nebula (see Fig. \ref{fig:AO}), the passivation timescale needs to be estimated. The number of particles on the surface of an agglomerate is given by
\begin{eqnarray}
N_{surf}&=&\frac{2 \pi r^2  l^{\frac23}}{2 \pi r_{\mu m}^2}\mbox{,}
\end{eqnarray}
where $r_{\mu m}$ is the radius of a monomer. The rate at which monomer particles impact on a dust aggregate's surface is given by
\begin{eqnarray}
\frac{dN_{coll}}{\dif t}&=&\rho_{0} r^2 \pi v_{imp} f \mbox{.}
\end{eqnarray}
We found in our experiments (e.g. Fig. \ref{fig:erhist}) that an agglomerate is passivated after about 10 times the number of surface monomers have impacted on the agglomerated. Thus, the passivation time is
\begin{eqnarray}
t_{pass}&=&10 N_{surf} / {\frac{dN_{coll}}{\dif t}} \mbox{.}
\end{eqnarray}
This passivation timescale $t_{pass}$ is smaller than a day for all reasonable parameters and orbital radii in the disk. Therefore, passivated agglomerates are dominating the erosion process so that we use the experimental erosion efficiency from the passivated dust samples (dot-dot-dashed line in Fig. \ref{fig:pas} and Eq. \ref{eq:fit}).

\subsubsection{Coagulation Timescale}\label{kap:CT}
For the coagulation timescales we use the results obtained by \citet{Nomura2006}.
Their model is based on the assumption that compact particles (with a filling factor of 100\%) that grow to a certain size are removed from the coagulation process by radial drift. In other models \citep[e.g.][]{Blum, Brauer2008, Zsometal2010}, the coagulation is counteracted by fragmentation and bouncing of colliding particles, which is not considered here. As macroscopic protoplanetary dust aggregates possess filling factors of about 40\%, we will adapt the model by \citet{Nomura2006} to our non-compact dust aggregates. The coagulation timescale of \citet{Nomura2006} is given by
\begin{eqnarray}
\tau_k =\frac{1}{\rho_{mi} A_ {mi} v_r},
\end{eqnarray}
with $\rho_m$, $A_ m$, and $v_r$ being the number density of dust aggregates with mass $m$, their collision cross sections, and the relative velocities, respectively.

The coagulation timescale $\tau_k$ depends on the filling factor $l$ via the cross section
\begin{eqnarray}
 A_ {m}=A_ {m1} l^{-\frac23},
\end{eqnarray}
which leads to
\begin{eqnarray}
\tau_k =l^{\frac23} \frac{1}{\rho_{m} A_ {m1} v_r}\mbox{,}
\end{eqnarray}
where $A_ {mi1}$ is the cross section of a particle in case $l=1$ and $\rho_{m}$ is the number density of these particles.
The relative velocity is approximated for simplicity by the velocity caused by turbulence, $v_r = v_{turb}$, and is assumed to be independent of the particle size and filling factor.


In \citet{Nomura2006} coagulation is balanced by the radial drift of the particles with a velocity
\begin{eqnarray}
v_{d} = g_d \tau_f \label{eq:vr}\mbox{,}
\end{eqnarray}
where  $g_d$ is the local acceleration caused by the sub-Keplerian motion of the particle due to gas drag.
The friction time of a given particle is
\begin{eqnarray}
\tau_f=\frac29 \frac{\rho_p}{\rho_g}\frac{r_d^2}{\nu_m}\mbox{,} \label{eq:tauf}
\end{eqnarray}
with $\rho_p$, $\rho_g$, $r_d$, and $\nu_m$ being the mass density of the dust particle (proportional to $l$), the mass density of the gas, the maximum radius particles can grow, and the mean molecular velocity of the gas, respectively \citep[see][]{Weidi77}.
Here, $\rho_p$ depends on the bulk density $\rho_b$ of the dust monomers and the filling factor of the dust aggregate through
\begin{eqnarray}
\rho_p=l \rho_b .
\end{eqnarray}
The drift timescale is given by
\begin{eqnarray}
\tau_d=\frac{S_D}{v_{d}}\mbox{,}\label{eq:td}
\end{eqnarray}
where $S_D$ is an arbitrary drift length.
As the only dependency of Eqs. \ref{eq:td}, \ref{eq:tauf}, and \ref{eq:td} on the volume filling factor $l$ is in $\rho_p$, $\tau_d$ is thus proportional to $l^{-1}$. This means that the coagulation timescale scales with $l^{\frac23}$, whereas the drift timescale scales with $l^{-1}$.

To use the results of \citet{Nomura2006}, a new maximum particle radius $r_d$ (dependent on  $l$) has to be found with which coagulation and radial drift still balance each other.
The ratio of the two timescales,
\begin{eqnarray}
\frac{\tau_k(l)}{\tau_d(l)}=\frac{2 l^{\frac53} g_d \rho_b r_d^2}{9 {\rho_{m} A_{m1}} v_{turb} \rho_g \nu_m S_d} ,
\end{eqnarray}
is proportional to $l^{\frac53}$ and proportional to $r_d^2$. This means the coagulation with a filling factor of $l$ is still balanced by the radial drift if $r_d$ is multiplied by a factor $l^{-\frac56}$. Therefore, the maximum particle radius at 1AU for a volume filling factor of  $l = 0.40$ is about 0.5 m. 
From these considerations, it follows that the results of \citet{Nomura2006} are still valid for our lower filling factors if the maximum size of their particles is increased by a factor $l^{-\frac56}$. The original coagulation timescale will then be changed by $l^{-\frac23}$. However, the relative size distribution derived by \citet{Nomura2006} will remain unchanged. The original coagulation timescales are found in \citet{Nomura2006} (their Fig. 5), which is the time at which 70\% of the total dust mass is in particles with a critical size at which they are lost from the coagulation process by fast radial drift. The dependency of the coagulation timescale on the disk radius is $\tau_k = 70$ years $\times ~ l^{\frac23}$ for $\alpha=10^{-2}$ and $\tau_k = 500$ years $\times ~ l^{\frac23}$ for $\alpha=10^{-4}$, respectively. We interpolated the timescales for the other radii assuming a power law. The resulting coagulation timescale as a function of the Kepler radius is shown in Figs. \ref{fig:ET-2} and \ref{fig:ET-4} as solid lines in the top graph.

\subsubsection{Timescale of Radial Turbulent Transport of Micrometer-Sized Dust}
The  timescale of turbulent transport $\tau_{turb}$ of micrometer-sized dust is
\begin{eqnarray}
\tau_{turb}=\frac{R^2}{\nu_{turb}}.
\end{eqnarray}
The turbulent gas viscosity is given by 
\begin{eqnarray}
\nu_{turb}=\alpha c_s H,
\end{eqnarray}
\citet{Shakura}, with the sound speed
\begin{eqnarray}
 c_s=\sqrt{\frac{\gamma  k_B T}{m_{g}}},
\end{eqnarray}
where $\gamma$ is the adiabatic exponent of $H_2$, $k_B$ is Boltzmann's constant, and $m_g$ is the mass of an $H_2$ gas molecule. The temperature is assumed to be dependent on the Kepler Radius $R_K$ via
\begin{eqnarray}
T = 280 ~{\rm K} \left(\frac{R_K}{1 AU}\right)^{-\frac12}
\end{eqnarray}
\citep[see][]{SchrHen04}. The timescale of turbulent particle transport as a function of the Kepler radius for $\alpha=10^{-2}$ and $\alpha=10^{-4}$ is shown in Figs. \ref{fig:ET-2} and \ref{fig:ET-4}, respectively (top graphs, dotted line). It is much longer than the coagulation and the erosion timescale. Therefore the effect of turbulent particle transport is negligible for our considerations.

\subsubsection{Particle Size Limit Caused by Radial Drift}
The model of \citet{Nomura2006} is based on the assumption that {\it compact} particles that grow to a certain size are removed from the coagulation process by radial drift. \citet{Nomura2006} find in their Eq. 24 and the sentence below  that these particles can grow close to the midplane to 32 cm, 80 cm, and 6 cm at 1 AU, 5.2 AU, and 30 AU, respectively for a turbulence strength of $\alpha=10^{-2}$ and  $\alpha=10^{-4}$. The assumption of compact particles should now be dropped in favor of a realistic assessment of the porosity of the dust aggregates. We have seen above that fluffy dust aggregates get compacted by the continuous bombardment of micrometer-sized dust particles. Our model for the erosion efficiency (see Fig. \ref{fig:pas}), which assumes a mean coordination number of 4, fits reasonably well. Such a coordination number is related to a volume filling factor by the relation shown in \citet{vandelagemaatetal2001}. From their Fig. 4, we can conclude that the corresponding (surface) filling factor of the aggregates was increased to $l \approx 0.4$. To a similar result for the (bulk) filling factor, we get if we follow \citet{Zsometal2010} who calculate that macroscopic dust aggregates should have volume filling factors $l \approx 0.4$, due to low-velocity bouncing collisions. Thus, we assume in the following a volume filling factor of $l = 0.4$ for large dust aggregates in PPDs. For this, we transformed the dust-aggregate sizes of \citet{Nomura2006} with the factor $l^{-\frac56}$, calculated in Sect. \ref{kap:CT}. These transformed sizes are plotted as triangles in Figs. \ref {fig:ET-2} and \ref{fig:ET-4} (lower graphs). In between the data points, we assumed power laws, which are shown as solid lines in the lower graphs of Figs. \ref{fig:ET-2} and \ref{fig:ET-4}.

\subsubsection{Particle Size Limit Caused by Erosion}
In \citet{Nomura2006}, the coagulation of particles is limited due to the fact that large particles are lost by the drift toward the central star. However, in our case and at Kepler radii $R <$ 0.2 AU ($\alpha=10^{-2}$) and $R <$ 0.3 AU ($\alpha=10^{-4}$), the erosion timescale is much shorter than the coagulation timescale (see Figs. \ref{fig:ET-2} and \ref{fig:ET-4}, top graphs). Therefore, the particles can only grow until the onset of erosion at drift velocities of 26 $\rm m~s^{-1}$ (see Fig. \ref{fig:pas}). The dependency of the impact velocity, $v_{imp}$, on the particle radius is estimated using \citet{Weidi77} (his Fig. 3) for a particle density of 1000kg m$^{-3}$. The impact velocity can be approximated by
\begin{eqnarray}
v_{imp}= 26\frac{\rm m}{\rm s}\left(\frac{r}{0.21 \rm m}\right)^{\frac75}\mbox{.}\label{eq:radv}
\end{eqnarray}


The dependency of the impact velocity, $v_{imp}$, on the Kepler radius is estimated using \citet{Weidi77} (his Fig. 5). We have to cover a particle range from  0.08 m to 0.2m of our dust aggregates with a volume filling factor of 0.4. As \citet{Weidi77} assumed {\it compact} particles, we interpolated the curves in his Fig. 5 with our filling factor of $l = 0.4$, having in mind that the independent variable is the friction time of the dust aggregates. Therefore we used his 10-cm curve and approximate it with
\begin{eqnarray}
v_{imp}= 26\frac{m}{s} \left(\frac{R[AU]}{1.1 AU}\right)^{-\frac{3}{5}}\mbox{.}\label{eq:AUv}
\end{eqnarray}
To find  the particle radius for a given Kepler radius at which the drift velocity exceeds 26 m s$^{-1}$ and therefore causes erosion, we combine Eqs.~(\ref{eq:radv}) and (\ref{eq:AUv}) to
\begin{eqnarray}
v_{imp}= 26\frac{m}{s} \left(\frac{R[AU]}{1.1 AU}\right)^{-\frac{3}{5}}\left(\frac{r[m]}{0.21 m}\right)^\frac{7}{5},
\end{eqnarray}
set $v_{imp}$ to 26 m s$^{-1}$ and achieve
\begin{eqnarray}
r=0.21 {\rm ~m} \left(\frac{R[{\rm ~AU}]}{1.1 {\rm ~AU}}\right)^{\frac{3}{7}}\mbox{.}\label{eq:rAU}
\end{eqnarray}
Eqn.~(\ref{eq:rAU}) is given as a dashed dotted line in the lower graphs of Figs. \ref{fig:ET-2} and \ref{fig:ET-4}. The particles size above would be reduced further in case the impact velocity driven by turbulence which is neglected here would be taken into account.

\subsubsection{Number Density of Micrometer-Sized Particles}
At Kepler radii larger than 0.2 AU (for $\alpha=10^{-2}$) and 0.3 AU (for $\alpha=10^{-4}$), the timescale of the erosion is much larger than the coagulation timescale (see top graph in Figs. \ref{fig:ET-2} and \ref{fig:ET-4}). This means that particles can grow to the size predicted in Sect. \ref {kap:CT} and the number density of the micrometer-sized particles is given by Eq. \ref{eq:rhor}. For Kepler radii smaller than the above values, the timescale for the erosion of large particles is much shorter than the coagulation timescale. This means that particles can only grow as long as their relative motion to the gas is smaller than the threshold at which the erosion process sets in. To calculate the number density of micrometer-sized dust particles produced by the erosion process at Kepler radii lower than given above, it is assumed that for all particles, which coagulate to a size where the relative motion to the gas causes erosion, the further coagulation is compensated by erosion. To get the number density of micrometer-sized particles, the production by erosion has to be balanced by their re-coagulation.

In the following, $\rho_{00}$ denotes the total number density at a given Kepler radius if all the dust mass of the disk was in micrometer-sized particles and $\tau_{co}$ is the coagulation timescale to the particle size, r$_{max}$, at which erosion sets in. We assumed for simplicity that this is the same rate as the coagulation to dm-sized particles given in \citet{Nomura2006} their Fig. 5.   However, coagulation continues, but all the mass gain of such dust aggregates caused by coagulation is immediately eroded away in regions where the erosional timescale is faster than the coagulation timescale. Thus, we get as the source term for monomer dust grains
\begin{eqnarray}
\frac{\dif \rho_p}{\dif t}&=&\frac{\rho_{00}}{\tau_{co}} \mbox{,}  \label{eq:production}
\end{eqnarray}
which gives the production rate of micrometer-sized dust particles by erosion. This rate is balanced by the coagulation of the micrometer-sized particles into dimers, i.e.
\begin{eqnarray}
\frac{\dif \rho_p}{\dif t}&=&-\frac{\sigma v_B \rho_p^2}{\sqrt{2}}   \label{eq:destruction} \mbox{,}
\end{eqnarray}
where Eq. \ref{eq:siro} was used to integrate the coagulation, which takes place over the whole disk height. Combining Eqs. \ref{eq:production} and \ref{eq:destruction} leads to the stationary solution
\begin{eqnarray}
\rho_{ps}&=&\sqrt{\frac{\rho_{00}\sqrt{2}}{\tau_{co} \sigma v_B}}~. \label{eq:roso}
\end{eqnarray}
The number density of the micrometer-sized dust particles, resulting from erosion (source term) and coagulation (sink term), is shown in Fig. \ref{fig:ET-2} for $\alpha=10^{-2}$ and in Fig. \ref{fig:ET-4} for $\alpha=10^{-4}$ (second graphs), as a function of distance to the central star, respectively. Here, the solid line shows the region in which the erosion timescales is larger than the coagulation timescale and the dashed-dotted line shows where the erosion timescale is smaller than the coagulation timescale (see top graphs). We integrated all $\rm \mu$m-sized dust particles over the entire disk volume and got for their total mass $1.7 \times 10^{24} {\rm ~kg}$ for $\alpha=10^{-2}$ and $1.1\times 10^{24} {\rm ~kg}$ for $\alpha=10^{-4}$, respectively. Applying a minimum disk mass of 0.0425M$_\odot$ for the gas mass and a total mass ratio between dust and gas of 0.01, we get a mass fraction of the $\rm \mu m$-sized dust particles with respect to the total dust mass in the disk of $2.0\times 10^{-3}$ for $\alpha=10^{-2}$ and $1.3\times 10^{-3}$ for $\alpha=10^{-4}$, respectively.

\subsection{Consequences for Protoplanetary Disks}

Figs. \ref {fig:ET-2} and \ref{fig:ET-4} (top graphs) show that the process of impact erosion takes place at orbital distances smaller than  5 AU. At orbital distances smaller than 0.2 AU (for $\alpha=10^{-2}$) and 0.3 AU (for $\alpha=10^{-4}$), respectively, the erosion timescale of the large dust aggregates becomes smaller than the coagulation timescales. Hence, the dust aggregates can only grow as long as their relative velocity to the gas (an, thus, to the small dust particles) is smaller than the  threshold velocity for erosion. Dust aggregates, which grew to larger sizes at larger orbital distances, radially drift inward from the outer disk and are then eroded to that size. The latter effect is neglected in the following, because the number of particles that drift in from further out is smaller than the particles that are locally produced by coagulation in that area. In the second graphs of Figs. \ref {fig:ET-2} and \ref{fig:ET-4}, the steady-state number density of micrometer-sized particles that are produced by the above effects and removed by coagulation are shown. At orbital distances in between $\sim$5 AU and 0.2 AU (for $\alpha=10^{-2}$) and $\sim$5 AU and 0.3 AU (for $\alpha=10^{-4}$), respectively, the erosion does not considerably affect the size of the large agglomerates (but still produces many $\rm \mu m$-sized dust grains). At orbital distances smaller than 0.2 AU (for  $\alpha=10^{-2}$) and 0.3 AU (for  $\alpha=10^{-4}$), the erosion balances the growth of the large dust aggregates by coagulation. Therefore, the number density increase of micrometer-sized dust particles towards smaller disk radii is not as steep as further out in the disk, because the drift velocity is reduced due to the smaller size of the largest particles. Here, radial diffusion has been neglected because its timescale is very long (see dotted lines in the top graphs of Figs. \ref {fig:ET-2} and \ref{fig:ET-4}). In the bottom graphs of Figs. \ref {fig:ET-2} and \ref{fig:ET-4}, the consequences of the erosion on the radii of the largest dust aggregates are shown. Note that 70\% of the total dust mass of the disk is concentrated in the largest particles \citep{Nomura2006}. At disk radii larger than 0.2 AU (for $\alpha=10^{-2}$) and 0.3 AU (for  $\alpha=10^{-4}$), respectively, the radii of the particles are nearly unaffected by erosion. The growth of dust aggregates at these radii is only limited by the inward migration of the particles. At smaller disk radii, the growth of the large dust aggregates is limited by erosion, which means that their sizes are reduced to less than 60\% (at 0.1 AU) of the size without erosion. 

In summary, the process of collisional erosion leads to a number density of micrometer-sized dust particles of up to  $10^7$ m$^{-3}$ and limits the maximum dust-aggregate size to $\sim 0.1$ m at orbital distances of 0.3 AU. 
The ratio of $\rm \mu$m-sized particles to the total dust mass at a given Kepler radius is shown in the third graph of Figs. \ref {fig:ET-2} and \ref{fig:ET-4} for $\alpha=10^{-2}$ and $\alpha=10^{-4}$, respectively. For a minimum-mass solar nebula model, we integrated over the whole disk volume and found a total-mass fraction of the $\rm \mu$-sized dust particles of $2.0\times 10^{-3}$ for $\alpha=10^{-2}$ and $1.3\times 10^{-3}$ for a $\alpha=10^{-4}$, respectively.

\subsection{Comparison with Observational Data}
To compare the amount of micrometer-sized particles found above with observational data, we use the work of \citet{Furlan2006}. Typical observations of \citet{Furlan2006} show dust depletion factors of 100 to 1,000 which means that in their observed disks $10^{-2}$ - $10^{-3}$ of the total dust mass still resides in micrometer-sized particles, whereas the remainder dust mass is hidden in large dust aggregates. It is noted that the disk models used for their data interpretation have a slightly shallower density slope with Kepler radius than in the minimum-mass disk used here, but we do not think that this has an important influence on the observed mass depletion factor. In our bimodal model, 2.0$\times 10^{-3}$ and 1.3$\times 10^{-3}$ of the total dust mass is in $\rm \mu$m-sized dust particles for $\alpha=10^{-2}$ and  $\alpha=10^{-4}$, respectively. 70\% of the total dust mass is contained in large dust aggregates with a filling factor of 40\%. This reproduces the observations of \citet{Furlan2006} very well.

In case the disks are already evolved and their gas density in the midplane at 1 AU has dropped to $10^{-9}$ kg m$^{-3}$, which is $10^{-3}$ of the value of a disk of minimum mass with an interstellar dust-to-gas mass ratio, the drift velocities of 5 mm-sized dust aggregates (having a filling factor of 40\%), corresponding to the drift velocities of 2.7 mm compact particles (see above), are larger than 26 m s$^{-1}$ (see \citep{Weidi77}, Fig. 4) and enable erosion much more efficiently. In this case, no decimeter- or meter-sized dust aggregates are necessary to explain the production and existence of micrometer-sized dust grains. This could also explain the cutoff at millimeter-sized grains seen in many disks of \citet{Ricci2010}. In our model, the $\rm \mu$m-sized dust particles are produced at Kepler radii smaller than  $\sim$5 AU. 
Hence, our model predicts that the disk is depleted of micrometer-sized dust particles at large Kepler radii while somewhat larger grains are still present. 

\subsection{Irregular Dust Particles}\label{kap:irr}
As noted above, we used in our experiments monodisperse $\rm SiO_2$ spheres as simulated protoplanetary dust particles. In reality, protoplanetary dust consists of grains with all possible shapes and follows a size distribution. As we found in an earlier paper \citep{BlumSchr2006}, the tensile strength of dust aggregates consisting of polydisperse irregular $\rm SiO_2$ particles is about five times smaller than that of the particles we used in the experiment described in Sects. \ref{kap:EXACT} and \ref{kap:EIMER}. As we can assume that the particle-release energy is proportional to the tensile strength, the erosional threshold should therefore approximately occur at 0.5 times smaller velocities. This means that the erosional process should work with more realistic particles up to disk radii of about 10 AU (using the disk model by \citet{SchrHen04}). Hence, erosion should be properly treated in all simulations in which large dust aggregates (exceeding $\sim$10 cm in size) play a role and for the entire range of terrestrial-planet and the inner part of the giant-planet formation zone.


\section{SUMMARY}\label{kap:COCON}
In protoplanetary disks, realistic dust growth models \citep[e.g.][]{Zsometal2010}, which do not use a dust retention mechanism \citep{Okuzumi2009, Okuzumi2010a, Okuzumi2010b}, fail to explain the observed large amount of micrometer-sized particles \citep{DullemondDominik2005}. Because high relative velocities between meter-sized dust aggregates and micrometer-sized dust particles exist in protoplanetary disks, we experimentally investigated in a first step whether collisions at these relative velocities lead to an erosion of the meter-sized aggregates, which could, in turn, be a source of micrometer-sized dust particles. We found that, after an initial stage in which an impacting particle can erode up to 10 dust grains from the fluffy dust agglomerate, the agglomerate is surface-compacted and passivated against further erosion at velocities lower than 26 ms$^{-1}$. At larger velocities, the erosion is reduced to 10\% of its initial strength. In a second step, we explained and confirmed the occurrence of impact-induced erosion with a numerical model. In a third step, we developed an analytical disk model and implemented the experimentally found erosive effect. This model shows that erosion is a strong source of micrometer-sized dust particles in protoplanetary disks. The steady-state solution of the model shows that the micrometer-sized dust particles encompass a fraction of 0.20\% (for $\alpha$ of $10^{-2}$) and 0.13\% (for $10^{-4}$) of the total dust mass in a disk if we assume that 70\% of the disk mass is in the form of meter-sized dust aggregates with a filling factor of 40\%. A further outcome of our model is that larger dust aggregates get eroded very fast at small Kepler radii, because they possess higher relative velocities to the micrometer-sized particles. As their relative velocities are also dependent on their size, their growth is limited by erosion to $\sim$0.1 m at orbital distances smaller than a few tenth of an AU. Finally, we used our analytical model to discuss the amount of micrometer-sized particles in the observational data of \citet{Furlan2006}. We can reproduce their outcome with our erosion model. Assuming that the observed disks are already evolved and their gas density has dropped to $10^{-3}$ of a disk of minimum mass, drift velocities of 5-mm-sized dust aggregates (with 40\% filling factor) are large enough to enable erosion. In this case, no decimeter- or meter-sized dust aggregates are necessary to explain the production and existence of micrometer-sized dust grains. The erosive radius loss of grains larger than 5 mm could also explain the cutoff at millimeter-sized grains seen in many disks of \citet{Ricci2010}.

\bigskip
\bigskip

This research was supported by DLR grant 50WM0636 and 50WM0936.

\bibliography{ms}

\begin{thebibliography}{38}
\expandafter\ifx\csname natexlab\endcsname\relax\def\natexlab#1{#1}\fi

\bibitem[{{Blum}(2004)}]{Blum}
{Blum}, J. 2004, in Astronomical Society of the Pacific Conference Series, Vol.
  309, Astrophysics of Dust, ed. {A.~N.~Witt, G.~C.~Clayton, \& B.~T.~Draine},
  369--+

\bibitem[{{Blum} \& {Schr{\"a}pler}(2004)}]{BlumSchr2004}
{Blum}, J. \& {Schr{\"a}pler}, R. 2004, Physical Review Letters, 93, 115503

\bibitem[{{Blum} {et~al.}(2006){Blum}, {Schr{\"a}pler}, {Davidsson}, \&
  {Trigo-Rodr{\'{\i}}guez}}]{BlumSchr2006}
{Blum}, J., {Schr{\"a}pler}, R., {Davidsson}, B.~J.~R., \&
  {Trigo-Rodr{\'{\i}}guez}, J.~M. 2006, \apj, 652, 1768

\bibitem[{{Blum} \& {Wurm}(2000)}]{BlumWurm2000}
{Blum}, J. \& {Wurm}, G. 2000, Icarus, 143, 138

\bibitem[{{Blum} \& {Wurm}(2008)}]{BlumWurm2008}
---. 2008, Annu. Rev. Astron. Astrophys., 46, 21

\bibitem[{{Brauer} {et~al.}(2008){Brauer}, {Dullemond}, \&
  {Henning}}]{Brauer2008}
{Brauer}, F., {Dullemond}, C.~P., \& {Henning}, T. 2008, \aap, 480, 859

\bibitem[{{Chokshi} {et~al.}(1993){Chokshi}, {Tielens}, \&
  {Hollenbach}}]{Chokshi}
{Chokshi}, A., {Tielens}, A.~G.~G.~M., \& {Hollenbach}, D. 1993, \apj, 407, 806

\bibitem[{{Ciesla}(2007)}]{Ciesla2007}
{Ciesla}, F.~J. 2007, \apjl, 654, L159

\bibitem[{{Cuzzi} {et~al.}(1993){Cuzzi}, {Dobrovolskis}, \&
  {Champney}}]{Cuzzi1993}
{Cuzzi}, J.~N., {Dobrovolskis}, A.~R., \& {Champney}, J.~M. 1993, Icarus, 106,
  102

\bibitem[{{Dominik} \& {Tielens}(1997)}]{DominikTielens1997}
{Dominik}, C. \& {Tielens}, A.~G.~G.~M. 1997, \apj, 480, 647

\bibitem[{{Dullemond} \& {Dominik}(2005)}]{DullemondDominik2005}
{Dullemond}, C.~P. \& {Dominik}, C. 2005, \aap, 434, 971

\bibitem[{{Furlan} {et~al.}(2006){Furlan}, {Hartmann}, {Calvet}, {D'Alessio},
  {Franco-Hern{\'a}ndez}, {Forrest}, {Watson}, {Uchida}, {Sargent}, {Green},
  {Keller}, \& {Herter}}]{Furlan2006}
{Furlan}, E., {Hartmann}, L., {Calvet}, N., {et~al.} 2006, \apjs, 165, 568

\bibitem[{{Gammie}(1996)}]{Gammie1996}
{Gammie}, C.~F. 1996, \apj, 457, 355

\bibitem[{{G{\"u}ttler} {et~al.}(2010){G{\"u}ttler}, {Blum}, {Zsom}, {Ormel},
  \& {Dullemond}}]{Guettleretal2010}
{G{\"u}ttler}, C., {Blum}, J., {Zsom}, A., {Ormel}, C.~W., \& {Dullemond},
  C.~P. 2010, \aap, 513, A56+

\bibitem[{{Konstandopoulos}(2000)}]{Konsti1999}
{Konstandopoulos}, A.~G. 2000, Powder Technology, 109, 262

\bibitem[{{Lagemaat} {et~al.}(2001){Lagemaat}, {Benksteion}, \&
  {Frank}}]{vandelagemaatetal2001}
{Lagemaat}, J.~v.~d., {Benksteion}, K.~D., \& {Frank}, A.~J. 2001,
  J.~Phys.~Chem.~B, V, 105, 12434

\bibitem[{{Nakagawa} {et~al.}(1981){Nakagawa}, {Nakazawa}, \&
  {Hayashi}}]{Nakagawa1981}
{Nakagawa}, Y., {Nakazawa}, K., \& {Hayashi}, C. 1981, Icarus, 45, 517

\bibitem[{{Nakagawa} {et~al.}(1986){Nakagawa}, {Sekiya}, \&
  {Hayashi}}]{Nakagawa1986}
{Nakagawa}, Y., {Sekiya}, M., \& {Hayashi}, C. 1986, Icarus, 67, 375

\bibitem[{{Nomura} \& {Nakagawa}(2006)}]{Nomura2006}
{Nomura}, H. \& {Nakagawa}, Y. 2006, \apj, 640, 1099

\bibitem[{{Okuzumi}(2009)}]{Okuzumi2009}
{Okuzumi}, S. 2009, \apj, 698, 1122

\bibitem[{{Okuzumi} {et~al.}(2010{\natexlab{a}}){Okuzumi}, {Tanaka},
  {Takeuchi}, \& {Sakagami}}]{Okuzumi2010a}
{Okuzumi}, S., {Tanaka}, H., {Takeuchi}, T., \& {Sakagami}, M.
  2010{\natexlab{a}}, ArXiv e-prints

\bibitem[{{Okuzumi} {et~al.}(2010{\natexlab{b}}){Okuzumi}, {Tanaka},
  {Takeuchi}, \& {Sakagami}}]{Okuzumi2010b}
---. 2010{\natexlab{b}}, ArXiv e-prints

\bibitem[{{Poppe} {et~al.}(2000{\natexlab{a}}){Poppe}, {Blum}, \&
  {Henning}}]{PoppeBlumHenning2000}
{Poppe}, T., {Blum}, J., \& {Henning}, T. 2000{\natexlab{a}}, \apj, 533, 454

\bibitem[{{Poppe} {et~al.}(2000{\natexlab{b}}){Poppe}, {Blum}, \&
  {Henning}}]{PopBH}
---. 2000{\natexlab{b}}, \apj, 533, 472

\bibitem[{{Poppe} \& {Schr{\"a}pler}(2005)}]{PopSchr05}
{Poppe}, T. \& {Schr{\"a}pler}, R. 2005, \aap, 438, 1

\bibitem[{{Ricci} {et~al.}(2010){Ricci}, {Testi}, {Natta}, {Neri}, {Cabrit}, \&
  {Herczeg}}]{Ricci2010}
{Ricci}, L., {Testi}, L., {Natta}, A., {et~al.} 2010, \aap, 512, A15+

\bibitem[{{Schmitt} {et~al.}(1997){Schmitt}, {Henning}, \& {Mucha}}]{Schmitt}
{Schmitt}, W., {Henning}, T., \& {Mucha}, R. 1997, \aap, 325, 569

\bibitem[{{Schr{\"a}pler} \& {Henning}(2004)}]{SchrHen04}
{Schr{\"a}pler}, R. \& {Henning}, T. 2004, \apj, 614, 960

\bibitem[{{Shakura} \& {Sunyaev}(1973)}]{Shakura}
{Shakura}, N.~I. \& {Sunyaev}, R.~A. 1973, \aap, 24, 337

\bibitem[{{Stepinski}(1999)}]{Stepinski1999}
{Stepinski}, T.~F. 1999, in Lunar and Planetary Institute Conference Abstracts,
  1205--+

\bibitem[{{Wada} {et~al.}(2008){Wada}, {Tanaka}, {Suyama}, {Kimura}, \&
  {Yamamoto}}]{WadaII}
{Wada}, K., {Tanaka}, H., {Suyama}, T., {Kimura}, H., \& {Yamamoto}, T. 2008,
  \apj, 677, 1296

\bibitem[{{Weidenschilling}(1977)}]{Weidi77}
{Weidenschilling}, S.~J. 1977, \mnras, 180, 57

\bibitem[{{Weidenschilling}(1980)}]{Weidi1980}
---. 1980, Icarus, 44, 172

\bibitem[{{Weidenschilling}(1984)}]{Weidi1984}
---. 1984, Icarus, 60, 553

\bibitem[{{Weidenschilling}(1988)}]{Weidi1988}
---. 1988, {Formation processes and time scales for meteorite parent bodies}
  (Meteorites and the Early Solar System), 348--371

\bibitem[{{Weidenschilling}(2006)}]{Weidi2006}
---. 2006, Icarus, 181, 572

\bibitem[{{Weidenschilling} \& {Cuzzi}(1993)}]{WeiCu1993}
{Weidenschilling}, S.~J. \& {Cuzzi}, J.~N. 1993, in Protostars and Planets III,
  ed. E.~H. {Levy} \& J.~I. {Lunine}, 1031--1060

\bibitem[{{Zsom} {et~al.}(2010){Zsom}, {Ormel}, {G{\"u}ttler}, {Blum}, \&
  {Dullemond}}]{Zsometal2010}
{Zsom}, A., {Ormel}, C.~W., {G{\"u}ttler}, C., {Blum}, J., \& {Dullemond},
  C.~P. 2010, \aap, 513, A57+

\end{thebibliography}

\end{document}